\documentclass[twocolumn,showpacs,preprintnumbers,a4,prb]{revtex4}
\usepackage{graphicx}%
\usepackage{dcolumn}
\usepackage{amsmath}
\usepackage{amssymb}
\usepackage{bm}

\makeatletter
\def\btt#1{\texttt{\@backslashchar#1}}%
\DeclareRobustCommand\bblash{\btt{\@backslashchar}}%
\makeatother

\begin{document}

\preprint{de-a.tex}

\title{Phase diagram and magnetic properties of La$_{1-x}$Ca$_x$MnO$_3$ compound for $0\leq x \leq 0.23$.}

\author{M. Pissas and G. Papavassiliou}
\affiliation{Institute of Materials Science, NCSR,  Demokritos,
153 10 Aghia Paraskevi, Athens, Greece}

\date{\today}
\begin{abstract}
In this article a detailed study of La$_{1-x}$Ca$_x$MnO$_3$ (
$0\leq x \leq 0.23$) phase diagram using powder x-ray diffraction
and magnetization measurements is presented. Unfortunately, in the
related literature  no properly characterized samples have  been
used, with consequence the smearing of the real physics in this
complicated system. As the present results reveal, there are two
families of samples. The first family concerns samples prepared in
atmosphere ($P({\rm O}_2)=0.2$ Atm ) which are all ferromagnetic
with Curie temperature rising with $x$. The second family concerns
samples, where a post annealing in nearly zero oxygen partial
pressure is applied. These samples show a canted antiferromagnetic
structure for $0\leq x \leq 0.1$ below $T_N$, while for $0.125\leq
x <0.23$ an unconventional ferromagnetic insulated phase  is
present below $T_c$. The most important difference between
nonstoichiometric and stoichiometric samples concerning the
magnetic behavior, is the anisotropy in the exchange interactions,
in the stoichiometric samples putting forward the idea that a new
orbital ordered phase is responsible for the ferromagnetic
insulating regime in the La$_{1-x}$Ca$_x$MnO$_3$ compound.

\end{abstract}
\maketitle

\section{Introduction}
 Manganite perovskites
RE$_{1-x}$AR$_x$MnO$_3$ (RE=La and rear earths, AR=Ca, Sr, Ba)
display interesting and puzzling structural, magnetic and
transport properties. The reason for this is the close interplay
between charge, spin and lattice degrees of freedom. One of the
most popular detailed studied system, concerning the transport,
structural, and magnetic
properties\cite{wollan55,goudenough55,schiffer95,ramirez96,okuda00,biotteau01,pissas03,dagotto01}
is the Ca-based La$_{1-x}$Ca$_x$MnO$_3$ ( $0\leq x \leq 1$)
compound. In stoichiometric LaMnO$_3$ at $T_{\rm RO}$, a
structural rhombohedral (R) to orthorhombic (O) transition occurs.
In the O phase the MnO$_6$ octahedra attempt specific tilting
system due to the particular value of the tolerance factor. Mn
ions in an undistorted octahedral oxygen coordination have an
electronic structure $d^4=t_{2g}^3e_g^1$ (one electron on a doubly
degenerated $e_g$-orbital). Subsequently at $T_{\rm JT}$ LaMnO$_3$
is transformed from O to another orthorhombic structure (O$^/$).
In the O phase the three octahedral Mn–O bond lengths are almost
equal. The transition at $T_{\rm JT}$ originates from a
cooperative Jahn-Teller structural
transition\cite{kanamori60,kugel82} resulting in to anisotropic
Mn-O bond lengths, with the long bond ordered in a two sublattice
fashion\cite{lamno3} in the $ac$ plane. This particular ordering
of the long bonds in the basal $ac$-plane ($Pnma$ notation) has
been connected with the orbital ordering of the $d_{x^2}$ and
$d_{z^2}$ orbitals. Finally, at $T_{\rm N}$ the compound is
ordered antiferromagnetically attempting the so-called
antiferromagnetic structure (${\bf m}||a$-axis) ferromagnetic
layers coupled antiferromagnetically along $b$-axis). Such an
ordering is connected with the orbital ordering. The
antiferromagnetic interactions along $b$-axis have to do with the
three $t_{2g}$ electrons. The strong overlap between half filled
orbitals $d_{x^2}$ at site 1 and an empty one $d_{x^2-y^2}$ at
site 2 gives the ferromagnetic exchange\cite{kugel82} (at
$ac$-plane, in accordance with the Goodenough-Kanamori-Anderson
rules).

The substitution of La by Ca, i.e. doping the compound with
holelike charge carriers, induces drastic changes of the
structural and magnetic/tranport properties. In the doping level
$0.0<x\leq 0.1$ the long magnetically ordered state for $T<T_{c}$
is a canted antiferromagnetic structure (CAF) closely connected
with the A-antiferromagnetic structure of LaMnO$_3$ compound,
where the collinear magnetic moments in the $ac$-plane, are canted
in such a way producing a net ferromagnetic component along
$b$-axis\cite{moussa99}. The anisotropy of the Mn-O bond lengths
in the O$^/$ phase are still present, with a tendency for
reduction as $x$ increases. Based on the assumption that this
ordering of the long bonds is connected with the orbital ordering,
the orbital ordering is preserved, also in the CAF regime.

With further doping $0.125<x<0.23$ the CAF phase is transformed to
a ferromagnetic insulating phase in contradiction with the
conventional double and superexchange models.  to 80 K this phase
follows the O$^/$ structure\cite{aken03} with moderate anisotropic
Mn-O bond lengths. As the doping concentration $x$ increases, the
static JT distortion weakens progressively and the system becomes
metallic and ferromagnetic for $x>0.23$. It is believed that in
the absence of a cooperative effect in this regime, local JT
distortions persist\cite{lynn96,dai00,baca98} on short time and
length scales. These short-range correlations, together with the
electron correlations, would create the effective carrier mass
necessary for large magnetoresistance.

\begin{figure*}[htbp]\centering
\includegraphics[angle=0,width=0.9 \columnwidth]{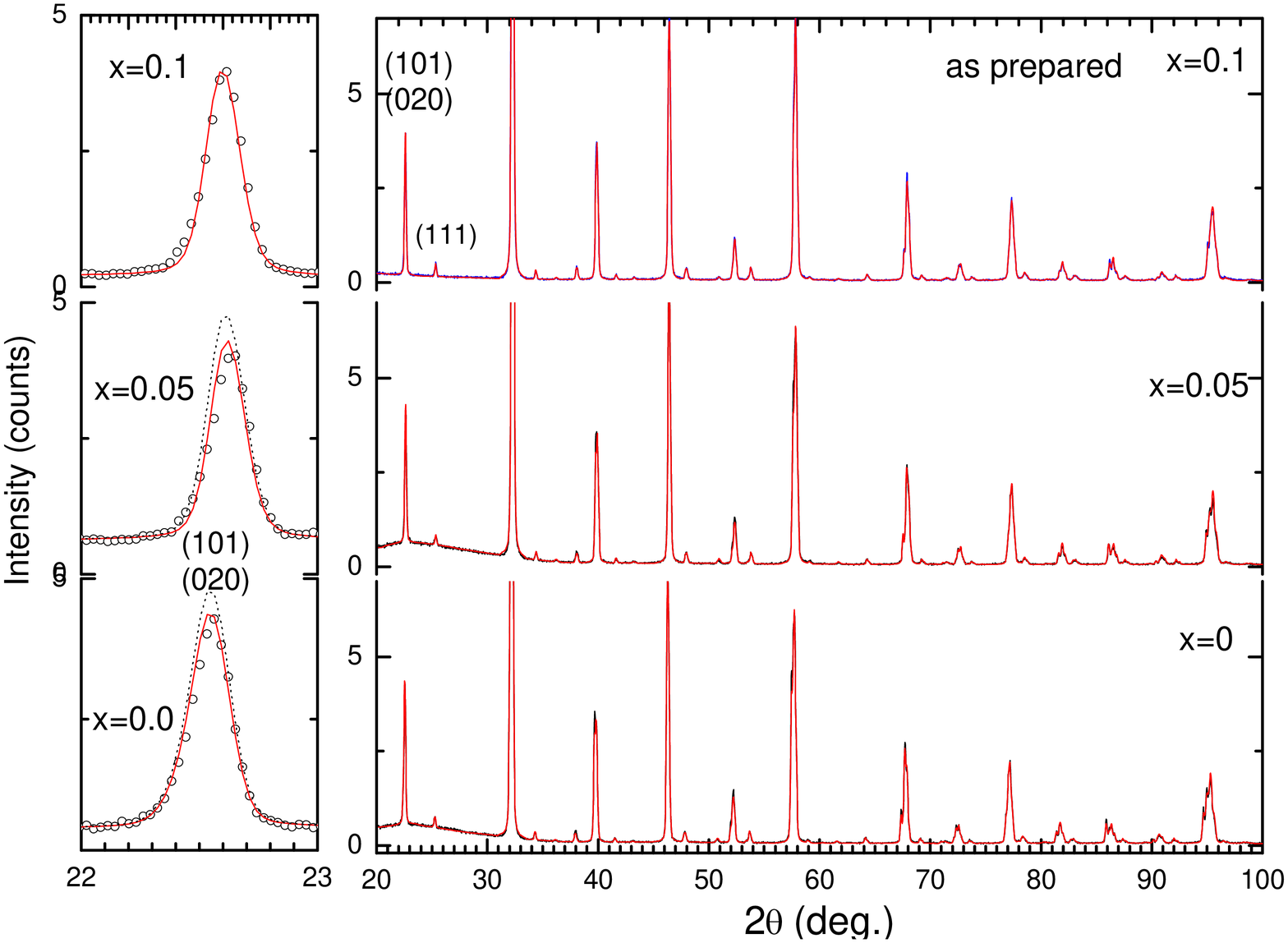}
\includegraphics[angle=0,width=0.9 \columnwidth]{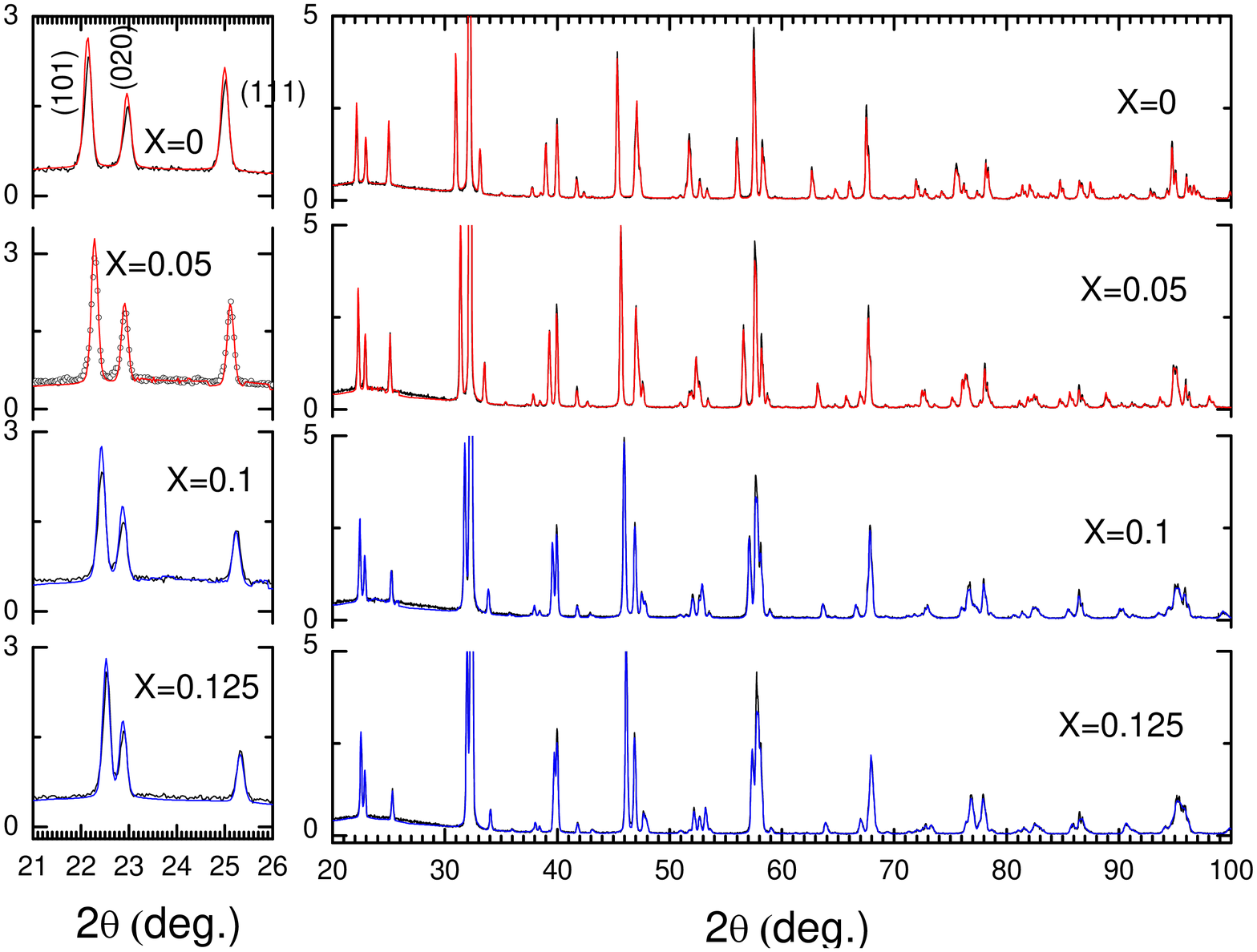}
\caption{XRD patterns of as prepared in air atmosphere at
$1400^{\rm o}$ (a) and post annealed  at $1000^{\rm o}$C in high
purity He La$_{1-x}$Ca$_x$MnO$_3$ samples (b). At the left of the
main panel is a magnification of the patterns near the (101),
(020) reflections. } \label{fig1}
\end{figure*}
The ferromagnetic insulating phase is one of the most puzzling
regime in the physics of magnanites perovskites. Several models
have been proposed to interpret it especially for the
La$_{1-x}$Sr$_x$MnO$_3$ compound, including new orbital and charge
ordering\cite{endoh99,yamada00} states. Although, a lot of
experimental studies have been devoted for the ferromagnetic
insulating regime of the La$_{1-x}$Ca$_x$MnO$_3$ compound, in most
of them, significantly nonstoichiometric samples were used,
leading,(as it is pointed out by Dabrowski et al. Ref.
\cite{dabrowski99}) in contradictory results if one compares
results of different groups. The nonstoichiometric samples
although showing ferromagnetic insulating behavior for $x<0.2$,
most probably concern samples with cation vacancies where the
unknown orbital state of the ferromagnetic insulating regime, is
probably in a liquid state. In addition, the concentration
gradients in the singe crystals (when they are free of vacancies)
make difficult the direct comparison of the results between
different composition and research groups. The main reason for
this is the fact that the crystal chemistry and physical
properties of the low Ca-samples depend on the partial pressure of
the oxygen during the sample preparation. As the results of the
present study reveal, there are two families of samples. The first
family concerns samples prepared in air atmosphere ($P({\rm
O}_2)=0.2$ Atm) displaying ferromagnetic behavior with Curie
temperature which diminishes as $x$ goes to zero. The second
family concerns samples where a post annealing in nearly zero
oxygen partial pressure is applied, leading to a complicated phase
diagram. A first attempt in understanding how the preparation
conditions influence the phase diagram of La$_{1-x}$Ca$_x$MnO$_3$
( $0\leq x \leq 0.2$) has been performed by Dabrowski et al.
 and Huang et al., in  Refs. \cite{dabrowski99,huang97,huang98}.
In these studies the role of the oxygen partial pressure has been
recognized as the key parameter influencing the Mn$^{+4}$ amount,
in addition to the nominal calcium content.  Taking into account
these observations one must be cautious in adopting theoretical
models before the basic solid state chemistry and the complete
characterization of the La$_{1-x}$Ca$_x$MnO$_3$ compound is made.

In the present paper we carry out a detailed study in respect of
bulk magnetic properties of La$_{1-x}$Ca$_x$MnO$_3$ ($0\leq x \leq
0.23$) in two sets of samples, in an effort, firstly to elucidate
the complications that originate from the nonstoichiometry of the
samples and secondly, using magnetic and resistivity measurements,
to understand the differences in magnetic properties between the
two families of compounds.
\section{Experimental details}
La$_{1-x}$Ca$_x$MnO$_3$ samples were prepared  by thoroughly
mixing stoichiometric amounts preheated La$_2$O$_3$, CaCO$_3$ and
MnO$_2$ following a solid-state reaction method, in air at
$1400^{\rm o}$C. We call these samples "air prepared" (AP). Part
of the AP samples was subsequently post annealed at $1000^{\rm
o}$C in high purity He flow. These samples are called reduced
samples (R). X-ray powder diffraction (XRD) data were collected
with a D500 SIEMENS diffractometer, using CuK$\alpha $ radiation.
The Rietveld refinement of the XRD data is performed by using the
FULLPROF program.\cite{carvajal92} DC magnetization measurements
were performed in a superconducting quantum interference device
(SQUID) magnetometer (Quantum Design MPMS2). For ac susceptibility
measurements a home made susceptometer was employed. The dc, the
drive and pickup coils are inside the cryogenic fluid, in order to
have measurements in a constant background.
\begin{figure}[h]\centering
\includegraphics[angle=0,width=0.9 \columnwidth]{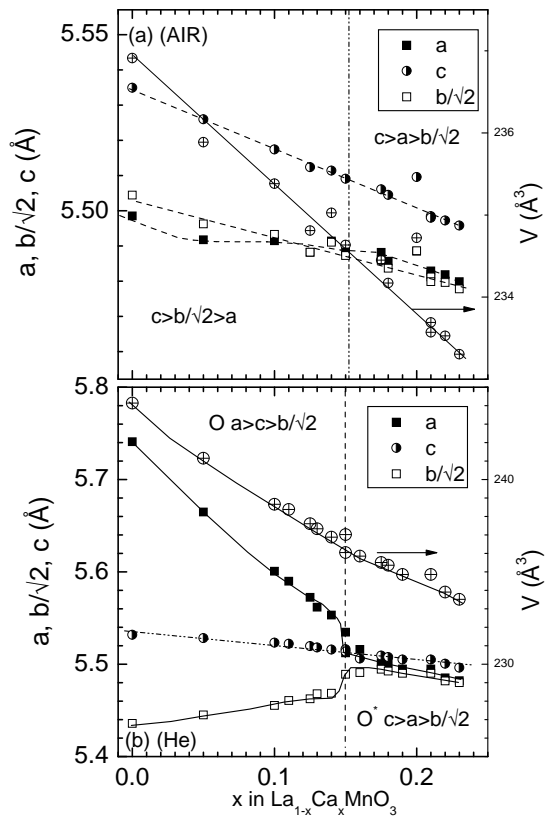}
\caption{(a) Concentrations dependance of unit cell constants and
volume for (AP) (a) and (R) (b) La$_{1-x}$Ca$_x$MnO$_3$ samples.}
\label{fig2}
\end{figure}
\section{x-ray diffraction data}
Figure \ref{fig1} shows part of the XRD patterns of the AP and R
samples. Based on Rietveld method both set of samples can
structurally be described by using the $Pnma$ space group. In
addition, the R samples display the cooperative Jahn-Teller
structure, which is characterized by the splitting of (101) and
(020) diffraction peaks. As the Rietveld refinement revealed, in
each MnO$_6$ octahedron there are three different bonds, with the
medium length Mn-O$_1$ bond (m) directed along the b-axis and the
long (l) and short (s) Mn-O$_2$ bonds alternating along the $a$
and $c$ axes ($Pnma$ notation).

Despite the fact that XRD data of all AP-samples are indicative
for single phase samples, all the samples (see below) are
ferromagnetic. No cooperative Janh-Teler distortion is present for
the AP samples as the unit cell constants and the Mn-O
bond-lengths point out, even for samples with $x<0.1$. {\it Most
probably, all these AP-samples concern cation deficient samples}.
One may claim that the oxygen content for this family of samples,
is larger than for the stoichiometric ones. However, the LaMnO$_3$
perovskite-type structure is a closed-packed LaO$_3$ lattice with
Mn ions in the octahedral sites completely surrounded by oxygens
ions. The possible sites for interstitial ions (tetrahedral and
octahedral) are surrounded by negatively charged ions (oxygens) as
well as by positively charged ions (lanthanum). It is very hard to
incorporate charged particles into these sites. Consequently, a
cation vacancies\cite{huang97} structure can explain the variation
of the crystal-chemistry and the physical properties of
La$_{1-x}$Ca$_x$MnO$_3$ system for $0\leq x \leq 0.23$. Most
importantly, refining the atomic positional and isotropic thermal
parameters, keeping the occupancy factors according to the nominal
composition, did not yield a completely satisfactory refinement
because some peaks (e.g. $(101), (020)$) did not have the adequate
intensity. Including as additional free parameters the occupation
factors for La and Mn, a clear improvement of the fit was
achieved, and moreover these peaks got the right intensity. The
results of the refinement clearly show that a nearly equal amount
of  La and Mn vacancies are present in the AP samples. The La and
Mn deficiency decreases as $x$ increases.

 Figure \ref{fig2} shows the concentration variation of the unit
cell parameters for AP and R samples. While for the AP samples the
cell constants are reduced monotonically as $x$ increases, the
corresponding variation of the R samples show a clear change in
the slope at around $x\approx 0.15$. For higher $x$ there is not
substantial difference between the variation of the cell constants
of the AP and R samples. It is interesting to be noted that the
unit-cell volume changes monotonously with $x$
 as is expected from the smaller ionic radius
both of the Ca$^{+2}$ and Mn$^{+4}$. However, the slope differ by
a factor of two in the two families of samples, e.g.
$(d(V/V_0)dx=-0.1455$ for AP samples and $(d(V/V_0)dx=-0.27$, for
the R samples, ($V_0$ is the unit cell volume at $x=0$ for AP and
R samples respectively). The difference in the unit-cell reduction
slope most probably is related with the cooperative Jahn-Teller
distortion, which is present only in the R samples. The abrupt
changes of the unit-cell parameters for $x<15$ is the result of
the intersection of the Jahn-Teller transition curve (e.g for
$x<0.15$, $T_{\rm JT}(x)>300$ K). Both families of samples exhibit
a GdFeO$_3$ distortion of the perovskite structure. Orthorhombic
perovskites have been separated into type O$^*$ ($b/c>\sqrt{2}$,
$Pnma$ notation), wherein the predominant distortion is octahedral
tilting as in GdFeO$_3$ and type O$^/$ ($b/c<\sqrt{2}$, wherein
the predominant distortion is driven by the cooperative
Jahn-Teller effect. The R samples at $T=300$ K up to $x=0.15$ are
described by the O$^/$-structure. We must note that the difficulty
in preparing homogenous samples in the $x=0.15$ regime is related
with the fact that the cooperative JT distortion, at the
particular concentration, occurs near $T=300$ K.

\begin{figure}[htbp]\centering
\includegraphics[angle=0,width=0.7 \columnwidth]{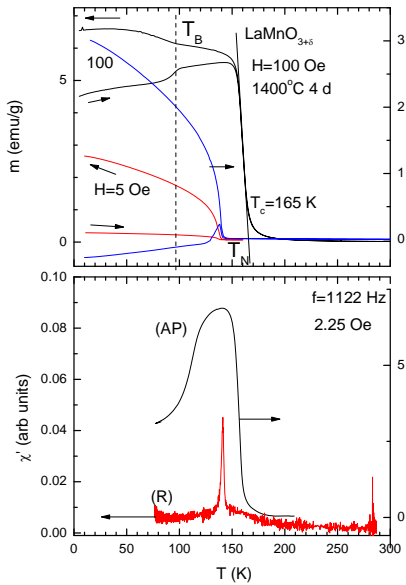}
\caption{(a) Temperature dependence of the dc-magnetic moment  and
 initial ac susceptibility for the air and He atmosphere prepared LaMnO$_3$ samples.
 The dc-magnetic moment was measured in ZFC and FC mode (see main
 text)
 } \label{fig3}
\end{figure}
\begin{figure}[htbp]\centering
\includegraphics[angle=0,width=0.7 \columnwidth]{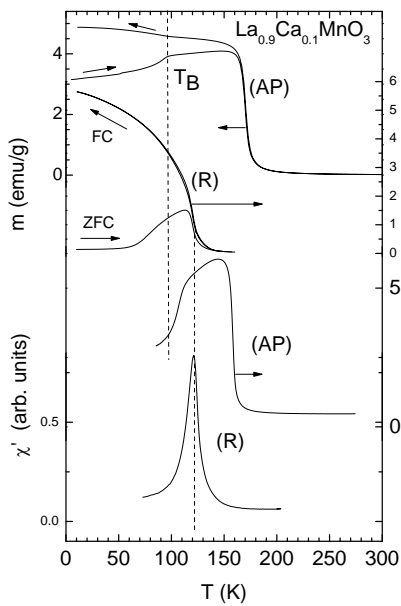}
\caption{(a) Temperature dependence of the dc-magnetic moment  and
 initial ac susceptibility for the air and He atmosphere prepared
 La$_{0.9}$Ca$_{0.1}$MnO$_3$ samples. The dc-magnetic moment was
 measured in ZFC and FC mode (see main
 text)
 } \label{fig4}
\end{figure}

\section{Magnetic measurements}
Proceeding further we carry out both dc- magnetic moment and
ac-susceptibility measurements in order to estimate the magnetic
phase diagram of the two families of samples. Figure \ref{fig3}
shows the temperature dependence of the magnetic moment for both
AP and R LaMnO$_3$ samples. The AP sample shows a paramagnetic to
ferromagnetic transition at $T_c=165$ K. The temperature variation
of the magnetic moment shows strong hysteretic behavior depending
on the measuring mode. In zero field cooling mode (ZFC) the moment
initially increases slightly up to 80 K. In a relatively narrow
temperature interval a step-like increasing of the magnetic moment
is observed. We characterize this feature by its onset
temperature, during cooling, denoted by $T_B$. Subsequently the
moment increases up to 150 K and then sharply decreases at Curie
temperature. The width of the transition at $T_c$ is relatively
sharp in comparison with ferromagnetic-paramagnetic transitions
observed in higher concentrations. This sharp decreasing of the
magnetic moment may be implying a first order magnetic transition.
However, the absence of measurable hysteresis in this regime is
not in favor of a clear first order transition. On cooling under
magnetic field the magnetic moment below 150 K shows a pronounced
hysteresis. This thermal irreversibility of magnetization is
observed below $T_{irr}$ being a common feature of all types of
magnetic systems showing magnetic hysteresis behavior. At $T_B$
the FC branch displays a slope change so that the moment for
$T<T_B$ rises more rapidly than above $T_B$. The particular shape
of the $m(T)$ curve appears in all AP samples for $0\leq x \leq
0.2$ (see below). The lower panel of Fig. \ref{fig3} shows the
real part of the ac-susceptibility for both the AP and R samples.
Let us firstly discuss, the data for AP sample. Below the Curie
temperature, $\chi'(T)$ increases down to the temperature where
the ZFC and FC branches of the dc-moment display hysteretic
behavior. Below this temperature the $\chi'(T)$ decreases rapidly,
and finally becomes nearly horizontal for $T<T_B$. The
corresponding imaginary part (not shown) shows two peaks: one at
the $T_c$ and the other at $T_B$. It is interesting to note that
the ac-susceptibility measurements do not show irreversible
behavior.

On the other hand, the R sample shows radically different
behavior. Both ac-susceptibility and the dc-magnetic moment show a
lower transition temperature $T_c=145$ K. The ZFC and FC branches
of the $m(T)$-curve show strong hysteretic behavior from the $T_c$
with the ZFC branch to be far below the FC one. Isothermal
magnetization measurements at $T=5$ K (not shown) revealed a
behavior typical of a canted antiferromagnet, in agreement with
neutron diffraction measurements\cite{biotteau01}(the neutron
diffraction data mainly show an A-type antiferromagnetic
structure, because the ferromagnetic component is very small). The
real part of the external susceptibility (uncorrected for
demagnetizing effects) $\chi'(T)$ displays a very narrow peak at
the transition temperature.

In a case of For a canted antiferromagnet the ac-susceptibility
along the ferromagnetic axis (e.g. the $b$-axis in our case), is
expected to diverge above and below $T_N$. On the other hand, the
ac susceptibility along the other orthogonal axes display the
usual behavior for a antiferromagnetic material. In our powder
sample the response is governed by the divergent part, which is
larger than the antiferromagnetic one. For a canted
antiferromagnet the susceptibility is expected to diverge on the
interval $\Delta T=(T_N-T_0)/T_N$ where $\Delta T=(D/\sum_i
|J_i|)^2$, $T_0$ is the temperature where the susceptibility
begins to diverge, and $D$ is the anisotropy constant of crystal
field type $DS_z^2$, or Dzyaloshinskii-Moriya interactions.
\cite{moriya} The sharp fall in $\chi'$ below $T_N$ can be
ascribed to the onset of coercivity. As the $x$ increases (e.g.
the samples 0.05 and 0.1) the magnetic measurements for both AP
and R-samples are practically similar with the LaMnO$_3$ sample.
{\it  It is interesting to note that the width of the
susceptibility peak increases with $x$. A fact which may be
related with a softening of $\sum_i |J_i|)^2$ with $x$}. In Figure
\ref{fig4} plotted are the dc-magnetic moment and
ac-susceptibility measurements for the $x=0.1$ samples in order
the differences with the $x=0.0$ sample to be emphasized. The
AP-prepared sample shows exactly the same behavior as the
corresponding $x=0.0$ sample except a slightly higher $T_c$. The
R-sample continues to show a behavior characteristic of a
canted-antiferromagnet.

\begin{figure}[htbp]\centering
\includegraphics[angle=0,width=0.7 \columnwidth]{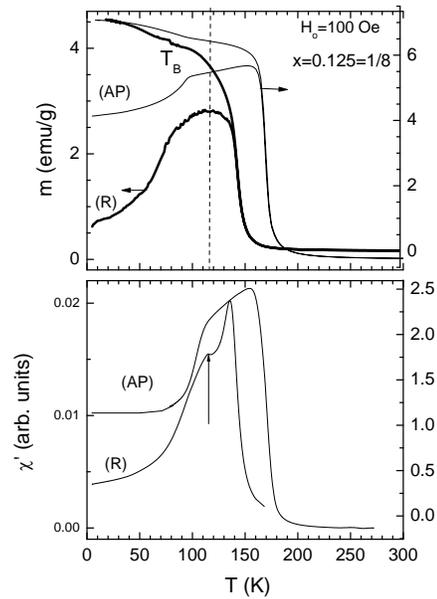}
 \caption{Temperature dependance of the dc-magnetization (a) and initial
 ac-susceptibility (b) for the AP and R La$_{0.875}$Ca$_{0.125}$MnO$_3$ samples, respectively.}
\label{fig5}
\end{figure}

\begin{figure}[htbp]\centering
\includegraphics[angle=0,width=0.7 \columnwidth]{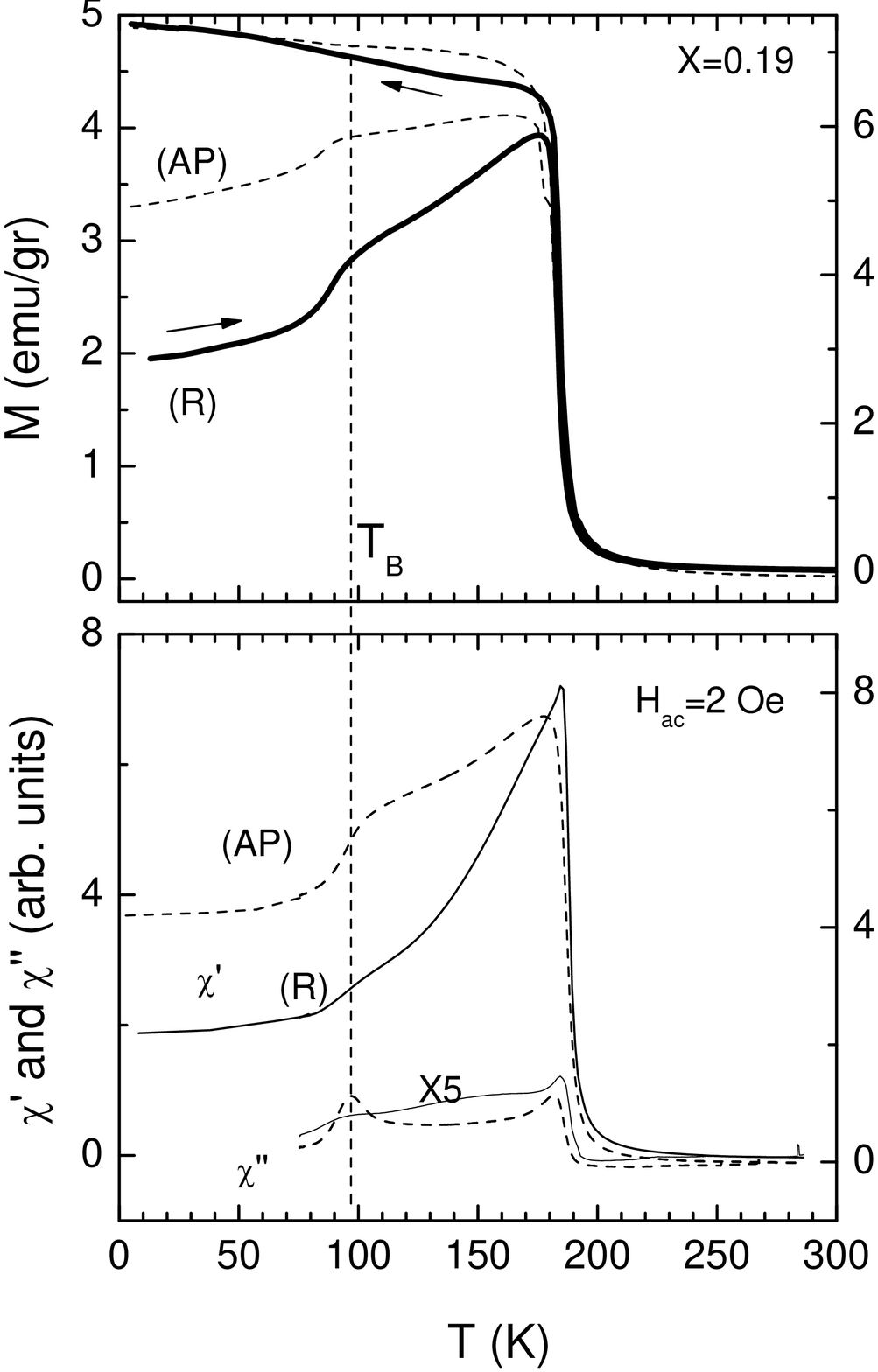}
 \caption{Temperature dependance of the dc-magnetization (a) and initial
 ac-susceptibility (b) for the AP and R La$_{0.81}$Ca$_{0.19}$MnO$_3$ samples, respectively.} \label{fig6}
\end{figure}

The magnetic measurements of the AP-sample $x=0.11$, shows similar
behavior as all the AP samples. The $\chi'$ of the R $x=0.11$
sample shows two features in place of the single peak observed for
$x\leq 0.1$. At the high temperature side of the peak of $\chi'$
appears a clear shoulder, indicating that this sample displays two
transitions. The transition located at the peak most probably is
related with CAF transition of the $x=0.1$ sample, while the
shoulder with a ferromagnetic transition. Similar behavior has
been observed by Biotteau et al.\cite{biotteau01} in neutron data
for a single crystal with $x=0.1$. Fig. \ref{fig5} shows the dc
magnetic moment and ac-susceptibility measurements for both AP and
R $x=0.125$ samples. The situation in the R $x=0.125$ sample is
more clear. Here, a ferromagnetic transition at $T_c\approx 150$ K
and a shoulder at 115 K (see arrow in Fig. \ref{fig5}) are
observed. We attribute this anomaly to antiferromagnetic
transition which in this sample occurs below $T_c$. The dc
magnetic moment displays hysteretic behavior bellow $T_{\rm
irr}\approx 140$ K, between of the ZFC and FC branches. The ZFC
branch although increases non linearly with temperature, does not
exhibit any other feature. At $T=115 {\rm K} \equiv T_N$ displays
a maximum.  The same behavior was observed in all the samples up
to $x=0.15$ concerning the peak at $T_c$ in the $\chi'$ of the R
samples. The shoulder now has been replaced by a broad shoulder.
For $T<80$ K, $\chi'$ becomes nearly horizontal. For $x>0.15$ the
dc magnetic measurements are not essentially different in respect
of the AP samples, except the lower $T_c$ for the R samples. The
step-like increment of the magnetic moment in the ZFC branches are
present in both families of samples. The ac-susceptibility
measurements for the R samples show an asymmetric peak at $T_c$
and a small reduction at the temperature region where the
step-like variation of the dc moment occurs. A representative
example is shown in Fig. \ref{fig6} for the $x=0.19$ sample. In
the R $x=0.19$ sample the dc-magnetic measurements show the
irreversibility at $T_{\rm irr}\approx 180 $ K and the step at
$T_B$. The corresponding $\chi'(T)$ shows an asymmetric peak at
$T_{\rm irr}$ then decreases and finally forms a shoulder at $T_B$
before becoming horizontal.

\begin{figure}[htbp]\centering
\includegraphics[angle=0,width=0.7 \columnwidth]{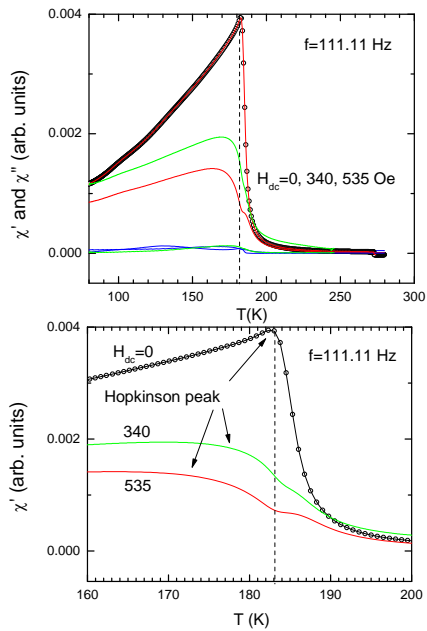}
 \caption{Temperature variation of the linear fundamental ac susceptibility for various dc fields of
 the $x=0.19$ R sample. }
 \label{hopkinson}
\end{figure}

The ac susceptibility for the samples with $x>0.14$ increases
rapidly as $T_c$ is approached from above, passing though a
maximum at a temperature somewhat below $T_c$. This maximum
originates from the Hopkinson effect\cite{williams} and not from
of critical effects. The Hopkinson effect is due to the rapid
increase in anisotropy with decreasing temperature below $T_c$,
particularly when its value begins to exceed the ac
field.\cite{chikazumi} The application of an external static
biasing field results in a rapid suppression in both amplitude and
temperature of this principal maximum. The measurements under a dc
field revealed a smaller secondary peak, which decreases in
amplitude and moves upward in temperature as the applied field
increases. This feature is a direct manifestation of critical
fluctuation in a system approaching a second order paramagnetic to
ferromagnetic transition and it is uniquely revealed by
susceptibility measurements. This behavior clearly appears  in
Fig. \ref{hopkinson} for the $x=0.19$, R sample. At this point we
would like to point out that a pronounced Hopkinson peak is absent
in the AP sample and in both families of samples for $x>0.2$.

Our magnetic measurements for the R $x=0.19$ sample are in good
agreement with the single crystal ones of Hong et al. ( Ref.
\onlinecite{hong02}). The anomaly at $T_B$ was attributed to
freezing of a cluster-glassy phase which appears bellow $T_c$.
According to their neutron diffraction data this glassy phase
concerns $0.4 \mu_B$ per Mn ion and has to do with charge/orbital
fluctuations. For $x>0.2$ both families of samples show typical
behavior of a low anisotropy soft ferromagnetic material.

\section{Electrical resistance measurements}
\begin{figure}[htbp]\centering
\includegraphics[angle=0,width=0.7 \columnwidth]{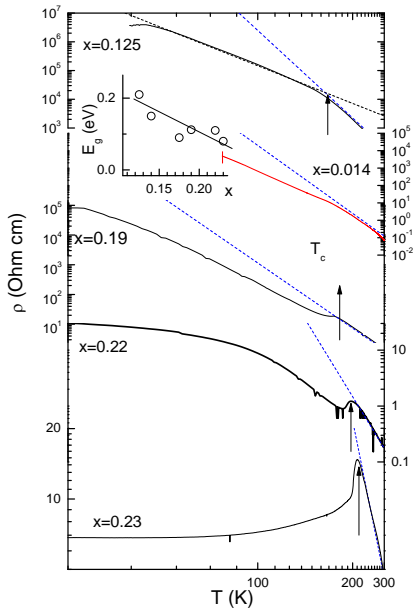}
 \caption{Temperature variation of the electrical resistivity of the R-samples
 in $\log-1/T$ plots. The straight dashed lines are plots of the equation ($\rho=\rho_0
\exp(E_g/kT)$. The estimated $E_g(x)$ is shown at the inset. }
 \label{res}
\end{figure}

Figure \ref{res} shows representative electrical resistance
measurements for R-samples in semi-logarithmic $1/T$ plots. This
kind of plots can show activation like ($\rho=\rho_0
\exp(E_g/kT)$, where $E_g$ is the activation energy) variation of
the resistivity. For $T>T_c$ our resistivity curves indicate
$E_g\sim 0.1-0.2$ eV, increasing as $x$ decreases (see inset of
Fig. \ref{res}). The $x=0.23$ R-sample displays insulating
behavior up to $T_c$ where an insulator to metal transition
occurs. For $x<0.23$ al the samples show insulating behavior. For
the $x=0.125$ R-sample, at $T_c$, the resistivity can be simulated
by two different activation energies above and below $T_c$, as it
is revealed from two linear segments in the $\log \rho$ vs. $1/T$
plot. At $T_c$ a reduction of the activation energy is observed.
As we approach the metal-insulator transition (near $x=0.23$) the
change of the activation energy occurs in a step like fashion,
while the resistivity curve bellow $T_c$ diverges from the
activation like behavior so that the $E_g$ effectively diminishes
as temperature decreases. The corresponding measurements for the
AP-samples also show insulating behavior with a slope change at
$T_c$ but the metallic behavior now occurs for lower $x=0.2$. We
must note that other researches have interpreted the step-like
change of the $\rho$ right below $T_c$ in the $\rho(T)$ curve for
the $x=0.22$ R-sample, as an insulator to metal behavior, which
subsequently is transformed to a metal to insulator behavior. To
our opinion this is a relative broad change of the electronic
structure at $T_c$, producing this abrupt change of $E_g$. In
other words, the ferromagnetic insulating regime is characterized
by ferromagnetic insulated ground state. The magnetic ordering
seems that abruptly renormalizes the electronic gap near the
center ($x=0.18$) of the ferromagnetic insulating regime. As the
low boundary $x=0.125$ is approached, this type of renormalization
becomes more smooth, while at the upper boundary the change is
more intense, since, there, we have complete change of the
electronic structure (metallic ground state).

\section{Phase diagram}

\begin{figure}[htbp]\centering
\includegraphics[angle=0,width=0.7 \columnwidth]{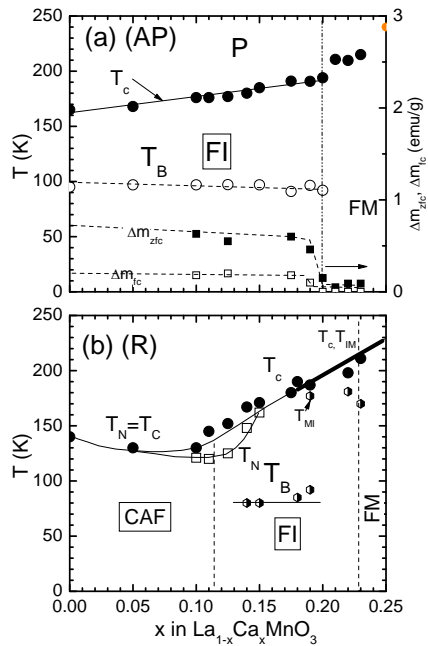}
\caption{(a) Phase diagram for the air atmosphere prepared
La$_{1-x}$Ca$_x$MnO$_3$ samples. The solid circles denote the
paramagnetic ferromagnetic transition. The open circles
corresponds to the onset temperature $T_B$ of the jump in the ZFC
dc-magnetization curves. The open and closed squares shows the
temperature variation of the magnetization jump-size during field
and zero-field cooling modes, respectively. (b) Phase diagram for
R (He atmosphere annealed) La$_{1-x}$Ca$_x$MnO$_3$ samples. The
solid circles denote the transition from the paramagnetic to
canted antiferromagnetic or ferromagnetic. The semi-filled circles
for $x>0.125$ correspond to the onset temperature $T_B$ of the
jump in the ZFC dc-magnetization measurements. } \label{fig7}
\end{figure}

\begin{figure}[htbp]\centering
\includegraphics[angle=0,width=0.7 \columnwidth]{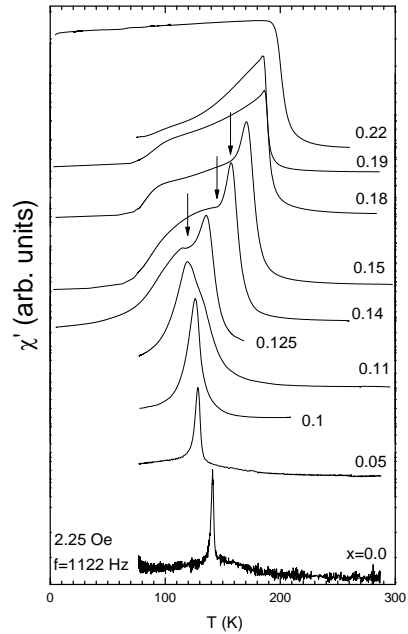}
 \caption{Temperature dependance of the $\chi'$ for the R La$_{1-x}$Ca$_{x}$MnO$_3$ samples
 ($0\leq x\leq 0.22$), respectively.} \label{fig8}
\end{figure}

Based on the magnetic measurements, the magnetic phase diagram for
both families of La$_{1-x}$Ca$_x$MnO$_3$ samples are determined
and plotted in Fig. \ref{fig7}. The AP samples display a
ferromagnetic transition at $T_c(x)$, which increases with $x$
(see Fig. \ref{fig7}(a)). In this figure plotted also are the
onset points $T_B(x)$, where the step increment of the magnetic
moment is observed. The temperature where this transition occurs
is nearly independent of $x$. Although this transition persists
and for $x>0.2$ the height of the magnetic moment "jump"
diminishes significantly there. The physical origin of the $T_B$
is the subject of several recent publications but its complete
elucidation has not yet been achieved.  Fig. \ref{fig7}(b) shows
the phase diagram of the R samples, which is radically different
in comparison with it of the AP samples, especially for $x<0.15$.
Our results for the R samples are in nice agreement with those of
Ref. \onlinecite{biotteau01,mandal03}, where stoichiometric single
crystals have been studied. For $0\leq x< 0.125$ the R samples
undergo a transition from a paramagnetic to a canted
antiferromagnetic state at $T_N$. The $x=0$ sample is an
$A_x$-type antiferromagnet with a small ferromagnetic component
along the $b-$axis ($Pnma$ notation). As $x$ increases the
ferromagnetic component of the canted magnetic structure increases
in expense of the antiferromagnetic one. For $x>0.08$ it is clear
that antiferromagnetic and ferromagnetic transitions occur at
different temperatures ($T_N<T_c$). For $0.125 \leq x\leq 0.2$
according to the magnetic saturation measurements all the samples
show ferromagnetic behavior. However, the real character of the
zero field ground state is not clear. The dc magnetic and
ac-susceptibility measurements show a rather complex behavior. Let
us describe  the possible physical origin of the special features
observed in magnetic measurements for R samples. Fig. \ref{fig8}
clearly shows this behavior. We have plotted the $\chi'(T)$ for
all the R-samples. The first transition appears as peak and has to
do with ferromagnetic interactions of the canted structure for
$0.125<x\leq0.15$. Subsequently, instead the peak decrease itself
considerably as in the CAF structure after a wide shoulder it
remains temperature independent down to zero temperature. At the
same time the dc-magnetic moment presents smaller difference
between ZFC and FC branches in comparison to the  CAF structure.
All these observations imply that we have a behavior where the
mechanism, which is responsible for the CAF structure  weakens
more and more as $x$ goes away from $x=0.125$. It seems that the
anomaly related with $T_N$ for $x<0.125$ is transformed to the
anomaly at $T_B$.

Having attributed the sharp peak right below $T_c$ in the middle
of the insulating ferromagnetic regime to the Hopkinson effect(
rapid increase of the magnetic anisotropy) we can conclude that in
the ferromagnetic insulating regime of the R-samples is related
with magnetic anisotropy increment. This increment reduces as the
metallic boundary is approached. On the other hand, this
anisotropy increases on approaching the canting antiferromagnetic
boundary. This significant conclusion may has to do with the idea
of some kind of orbital ordering\cite{hennion01,aken03} or orbital
domains\cite{papavassiliou03} occurring in the
ferromagnetic-insulating regime. A new kind of orbital ordering
below $T_c$, for a sample with nominal composition
La$_{0.85}$Ca$_{0.15}$MnO$_3$ has been proposed in Ref.
\onlinecite{lobanov00}. At $T=2$ K the high resolution neutron
diffraction pattern is compatible with a monoclinic distortion,
described from the space group $P2_1/c$. In this structural model
there are two nonequivalent MnO$_2$ layers alternating along the
$a-$axis leading to a specific pattern of Mn-O distances, implying
an unconventional orbital ordering type. This model may be
relevant with the ferromagnetic insulating state, where a specific
orbital ordering was proposed. However, the metallic resistivity
variation of the particular sample, takes in question the
generalization of this model in describing the FI phase.

We must note that, as the magnetic measurements show, in the
metallic regime ($x\geq 0.23$) no Hopkinson peak is present, a
fact which tells us that here the magnetic anisotropy goes to zero
smoothly at $T_c$. Concerning the AP-samples, they do not show the
Hopkinson peak also at $T_c$, implying that the mechanism which is
responsible does not operate in this case. Let us turn now on the
feature at the $T_B$. This feature is present in all the
AP-samples while it is less pronounced in the R-samples. It is
reasonable that in the R-samples the anomaly in the magnetic
measurements at low temperatures may be related with some kind of
antiferromagnetic interaction as the CAF boundary is approached.
These antiferromagnetic interactions may produce some kind of
glassiness with diminished weight as the metallic boundary is
approched.The detailed microscopic origin of the anomaly at $T_B$
is not clear.

In some works the feature at $T_B$ has been attributed to the
sudden change of the domain wail dynamic due to the domain wall
pinning effects.\cite{joy00} As also observed, small frequency
dependance of the peak in the $\chi''$, at temperature where the
step occurs, is not an indication of spin glass behavior. Many
magnetically ordered systems exhibit frequency dependance without
this to mean spin-glass behavior. This dependance has been
attributed to the freezing out of domain wall motion which is a
consequence of the rearrangements of electron states within the
domain wall. The low field ZFC magnetization follows the changes
of coercivity which is directly related with the
magnetocrystalline anisotropy. If this scenario is correct then
which is the origin of this sudden change in the
magnetocrystalline anisotropy? Laiho et al. \cite{laiho01} using
air atmosphere prepared samples based mainly on frequency
dependence of $\chi'$ at the region where the step was observed
they attributed this future to a reentrant spin glass phase. It is
well known that the compound with $x=0.5$ undergoes a first order
charge/orbital ordering antiferromagnetic transition below the
Curie temperature. This transition has as consequence the
appearance of a diminish of the magnetization in both FC and ZFC
magnetization branches. Something which is not observed in our
measurements, where ZFC and FC branches follow opposite direction
for $T<T_B$.

The physics of the AP samples near $x=0$ may be related with the
results of Algarabel et al. (Ref. \onlinecite{algarabel03}). In
this study, based on small angle neutron scattering data, was
suggested that the ferromagnetic phase (for $T<T_c$) comprises
from clusters increasing in size as the temperature decreases
reaching the value of 3-3.5 nm at low temperature. It is plausible
that the non-stoichiometric samples as the AP ones may consist
from regions rich in holes (then ferromagnetic) and poor in holes
(with glass behavior).

The magnetic measurements of Ref.
\onlinecite{markovich02a,markovich02b} practically coincide with
ours for the R-sample with nominal composition $x=0.15$. It is
most probable that in the single crystal the stoichiometry is
slightly lower from the nominal one. In the particular crystal
they observed a ferromagnetic to paramagnetic and metal-insulator
transition at $T_c\approx 180$ K, a ferromagnetic-insulator
transition at $T_{\rm FI}\approx 150$ K and a magnetic anomaly at
$T_B\approx 95$ K, which is related with a level off of the
resistivity.

Finally, it is interesting to compare the present magnetic
measurements with those of the La$_{1-x}$Sr$_x$MnO$_3$ at FI
regime occurring at $x=0.125$. This compound firstly exhibits a
cooperative Jahn-Teller first-order transition at $T_{\rm
JT}\approx 270 $ K, secondly, a transition towards a ferromagnetic
and metallic state, at $T_c=181$ K and then, a magneto-structural
first-order transition into a ferromagnetic insulating state, at
$T_B=159$ K. The transition at $T_B=159$ K is characterized by a
jump in the
magnetization\cite{dabrowski99,endoh99,uhlenbruck99,wagner00,liu01},
a typical for first-order transition delta function-like variation
of the specific heat, appearance superstructure peaks, significant
decreasing of the orthorhombicity and the three characteristic
Mn-O distances become very close to each other. In addition Moussa
et al. \cite{moussa03}have found a splitting of the spin waves, an
opening of a gap at ${\bf q}=(0,0,1/2)$ ( $Pnmb$ notation) and a
locking of the spin wave energy on the energy values of phonons.
All the above features occurring at $T<T_B$, are indicative for a
first order transition, most probably,  related with  a new
orbital order.\cite{endoh99}

Summarizing, in this work we have performed a systematic study of
La$_{1-x}$Ca$_x$MnO$_3$ ($0\leq x\leq 0.2$) with aim at advancing
the knowledge of the underlying mechanisms which influence their
structure and magnetic properties. The results of the present work
clearly demonstrate that the physical properties of the low doped
La$_{1-x}$Ca$_x$MnO$_3$ compound depend on the oxygen partial
pressure during the preparation influencing the Mn$^{+4}$ contain.
In order stoichiometric samples to be prepared a low oxygen
partial pressure is needed. The samples prepared in atmospheric
conditions for $x<0.16$ are cations deficient and in a such a way
so that the amount of the Mn$^{4+}$ to remain constant regardless
of $x$. The ferromagnetic insulating regime near the $x=0.2$
boundary for both families of samples is not homogeneous.

\end{document}